\documentclass[a4paper,12pt]{article}

\usepackage [T1]{fontenc}
\usepackage {calligra}
\usepackage [T1]{pbsi}
\usepackage{CJKutf8} 
\usepackage[colorlinks=true,linkcolor=blue]{hyperref}
\usepackage{amsmath,amssymb,amsfonts} 
\usepackage{graphicx}                 
\usepackage{color}                    
\usepackage{hyperref}                 
\usepackage{cite}
\definecolor{red}{rgb}{1,0,0}           
\definecolor{green}{rgb}{0,1,0}
\definecolor{blue}{rgb}{0,0,1}
\definecolor{darkblue}{rgb}{0,0,0.5}
\definecolor{lightblue}{rgb}{.5,.5,1}
\definecolor{lightgray}{gray}{.87}          
\definecolor{Dark}{gray}{.20}
\definecolor{pink}{rgb}{.95,0.82,0.92}  
\definecolor{yellow}{rgb}{1,1,0}
\definecolor{lightyellow}{rgb}{1,1,.5}
\definecolor{purple}{rgb}{0.7,0,0.85}
\definecolor{darkgreen}{rgb}{0,0.5,0}
\definecolor{orange}{rgb}{0.8,0.2,0.2}
\def \be {\begin{equation}}
\def \ee {\end{equation}}
\def \bea {\begin{eqnarray}}
\def \eea {\end{eqnarray}}
\def \ba {\begin{eqnarray}}
\def \ea {\end{eqnarray}}
\def \nn {\nonumber}

\def \rr {\raise.35ex\hbox{\small $\prime$}\kern-.17em{\mbox{\large $\imath$}}}
\def \del {\partial}
\def \dels {\partial\kern-.5em / \kern.5em}
\def \As {{A\kern-.5em / \kern.5em}}
\def \Ds {D\kern-.7em / \kern.5em}

\def \a {\alpha}

\def \b {\beta}

\def \d {\delta}
\def \eps {\epsilon}

\def \lam {\lambda}

\def \s {\sigma}

\def \om {\omega}

\def \z {(0)}
\def \zb {\bar{z}}

\def \vth {\vartheta}
\newcommand{\col}[2]{
\left[
#1 \atop #2
\right]
}
\def \Ac {{\cal A}}
\def \Bc {{\cal B}}

\usepackage[margin=1in]{geometry}

\newcommand{\solution}[1]{}

\newcommand{\beqs}{\begin{equation*}}
\def\beq{\begin{equation}}

\newcommand{\eeqs}{\end{equation*}}
\def\eeq{\end{equation}}

\newcommand{\beqas}{\begin{eqnarray*}}
\newcommand{\beqa}{\begin{eqnarray}}

\newcommand{\eeqas}{\end{eqnarray*}}
\newcommand{\eeqa}{\end{eqnarray}}

\begin{document}

\pagestyle{plain}

\setcounter{footnote}{0}
\setcounter{section}{0}

\begin{CJK}{UTF8}{bsmi} 

\begin{titlepage}
\begin{flushright}
UT-13-26
\end{flushright}
\begin{center}

\hfill
\vskip .2in

\textbf{\LARGE\centering 
Partition Function of Chiral Boson on 2-Torus \\
\vskip .6cm
from Floreanini-Jackiw Lagrangian
}

\vskip .5in
{\large \centering 
Wei-Ming Chen$\,{}^{a}$\footnote{e-mail address: tainist@gmail.com},
Pei-Ming Ho$\,{}^{a,b,c,}$\footnote{e-mail address: pmho@phys.ntu.edu.tw},
Hsien-chung Kao$\,^{d}$\footnote{e-mail address: hckao@phy.ntnu.edu.tw}, \\
Fech Scen Khoo$\,{}^{a}$\footnote{e-mail address: r00222076@ntu.edu.tw},
Yutaka Matsuo$\,{}^{e}$\footnote{e-mail address: matsuo@phys.s.u-tokyo.ac.jp}
\\

}
{\vskip 10mm \sl
${}^a$
Department of Physics and Center for Theoretical Sciences, \\
${}^b$
Center for Advanced Study in Theoretical Sciences, \\
${}^c$
National Center for Theoretical Sciences, \\
National Taiwan University, Taipei 106, Taiwan,
R.O.C. \\
${}^d$
Physics Department, 
National Taiwan Normal University, \\
Taipei 106, Taiwan,
R.O.C. \\
${}^e$
Department of Physics, Faculty of Science, University of Tokyo,\\
Hongo 7-3-1, Bunkyo-ku, Tokyo 113-0033, Japan
}\\
\vskip 3mm
\vspace{60pt}
\end{center}
\begin{abstract}

We revisit the problem of quantizing a chiral boson on a torus.
The conventional approach is to extract the partition function 
of a chiral boson from the path integral of a non-chiral boson.
Instead we compute it directly from the chiral boson Lagrangian of Floreanini and Jackiw
modified by topological terms involving an auxiliary field.
A careful analysis of the gauge-fixing condition for the extra gauge symmetry
reproduces the correct results for the free chiral boson, 
and has the advantage of being applicable to 
a wider class of interacting chiral boson theories.
\end{abstract}

\end{titlepage}

\setcounter{page}{0}
\setcounter{footnote}{0}

\section{Motivation and introduction}

\subsection{Motivation}

Folklore has it that there is no Lagrangian formulation 
for the quantum theory of chiral bosons.
In this work we examine this statement in detail for the special case 
of a chiral boson in two dimensions.

First of all,
there are various Lagrangian formulations for the classical theory of a free chiral boson 
\cite{Siegel:1983es,Floreanini:1987as,Henneaux:1988gg,Pasti:1996vs,Chen:2010jgb,HuangWH}.
These are well-defined theories useful for describing classical configurations.
Furthermore, at least for the free chiral boson theory,
there is no problem in using these Lagrangians for path integral quantization
on a base space with trivial topology.
The problem of Lagrangian formulation arises in the quantum theory 
only when both the base space and target space have nontrivial topology.

Let us consider the free field theory of a chiral boson in two-dimensional flat spacetime.
As a left-moving scalar field $\phi(\s_0+\s_1)$,
after Wick rotation ($\s_0 = i \s_2$),
the chiral boson becomes a holomorphic function $\phi(z)$ of the complex coordinate
\be
z \equiv \s_1 + i \s_2.
\ee
The problem of quantization arises when, for example,
the two-dimensional base space is compactified on a torus.
While there are many ways to define the Lagrangian of a chiral boson,
and for the uncompactified space they are equivalent,
the path integral for the torus depends on the choice of the Lagrangian
in a way that was not well understood,
and it was unclear whether any of them is correct.
On the other hand,
the partition function of a chiral boson can be deduced 
from that of a non-chiral boson via holomorphic decomposition,
or by finding a section on a holomorphic line bundle,
as we will review below.
The result is known to depend on the choice of a spin structure \cite{Witten:1996hc}.
Therefore,
if there exists a correct Lagrangian formulation for the chiral boson,
there has to be a way to introduce the choice of a spin structure in the Lagrangian.

With the choice of a spin structure,
the partition function of a chiral boson on a torus can be constructed 
from the theory of a non-chiral boson \cite{Witten:1996hc}.
Hence it may seem unnecessary to look for a Lagrangian formulation for the chiral boson.
However, 
there are many interesting chiral boson theories 
\cite{Oz:1990ia,Berkovits96,Bengtsson:1996ue,Perry:1996mk,
Pasti:1997gx,Bandos:1997ui,Dall'Agata:1998wh,Aganagic:1997zq,
Pasti:2009xc,Pasti:2012wv}
whose non-chiral versions are absent
(or at least it is unclear how to define them).
It will be useful to have a way of computing the partition function of a chiral boson 
directly from the Lagrangian.

Our goal is to understand how to modify a chiral boson Lagrangian 
so that it is suitable for path integral formulation.
As we do not want to spoil the local equation of motion for the chiral boson,
the modification is restricted to topological terms.
The guiding principle is to consider the most general topological terms 
consistent with the symmetries of the original theory.
When there are nontrivial cycles on the base space, 
the partition function is periodic in the zero mode of the source field.
The periodicity in the source field 
can be reduced to its fraction by taking a quotient.
This is the major property of a chiral boson Lagrangian that 
we explore to introduce spin structures into the action.

The plan of this paper is the following.
After a short review of related literature in Sec. \ref{review}
we revisit two approaches to derive the partition function 
of a free chiral boson in two dimensions from the path integral of a non-chiral boson 
in Sec. \ref{Holo-decomp} and Sec. \ref{Holo-bundle}.
Earlier results are generalized.
In Sec. \ref{Dir-calc}, 
we will demonstrate a way to derive the same result 
using the Lagrangian of Floreanini and Jackiw \cite{Floreanini:1987as} for a chiral boson
in the path integral formulation, 
with the addition of topological terms and an auxiliary field.
The ambiguity in the choice of a spin structure is encoded
in the choice of topological terms and the auxiliary field.
In Sec. \ref{s:BRS}, we give a gauge-fixed Lagrangian for 
the chiral boson model of Floreanini and Jackiw, 
and introduce a BRST symmetry.
It has the unique feature that ghost fields have dependence only on
one of the world-sheet variables (say the ``time" variable).
This feature reflects the nature
of the gauge symmetry of Lagrangian formulation of chiral boson
and removes undesirable degree of freedom on torus.
The conclusion is that, 
contrary to the folklore,
at least for the torus,
the Lagrangian formulation for the quantum theory of a chiral boson does exist.

\subsection{Introduction}
\label{review}

For the Lorentzian signature, 
the self-duality condition 
$d\phi = \ast d\phi$ is consistent in $4k+2$ dimensions.
The $(2k+1)$-form field strength $d\phi$ is defined in terms of 
a $2k$-form potential $\phi$ with the gauge transformation law
\be
\phi \rightarrow \phi' = \phi + d\lam,
\label{2k-form-gauge}
\ee
where $\lam$ is a $(2k-1)$-form.
For $k = 0$, 
the field $\phi$ is a scalar and 
there is nothing called $(-1)$-form $\lam$.
The gauge transformation by $d\lam$ 
is to be replaced by a shift of $\phi$ \cite{Siegel:1983es}
\be
\phi \rightarrow \phi' = \phi + c
\ee
for an arbitrary constant $c$.
This transformation does not change the physical state;
in other words, 
the constant mode of $\phi$ is not a physical observable.

Since the self-duality condition is a first order differential equation,
the Lagrangian formulation is nontrivial even at the classical level.
If we impose the self-duality condition as a constraint via a Lagrange multiplier,
the Lagrange multiplier acquires physical degrees of freedom 
following canonical formulation,
and so it is not the theory we want to study.
Siegel \cite{Siegel:1983es} realized that the problem 
can be avoided by imposing the square of the self-duality condition as the constraint,
and new gauge symmetries are introduced at the same time.
In fact, by introducing additional gauge symmetries in different ways,
there are many ways to write down a Lagrangian for chiral bosons in general dimensions
\cite{Floreanini:1987as,Henneaux:1988gg,Pasti:1996vs,Chen:2010jgb,HuangWH}.

The quantum theory of chiral bosons has also been studied in various aspects, 
including anomaly and bosonization 
\cite{Siegel:1983es,Eguchi-Ooguri,Verlinde:1986kw,Floreanini:1987as,Imbimbo:1987yt,
Dolgov:1987yp,Sonnenschein:1988ug,Bernstein:1988zd,Stone:1989cv,Harada:1989qp,McClain:1990sx,
Witten:1991mm,Wotzasek:1991un,Martin:1994np,Devecchi:1996cp,
Witten:1996hc,Dolan:1998qk,Henningson:1999dm,Witten:1999vg,Belov:2006jd,Giaccari:2008zx}.
The quantization of a chiral boson theory imposes new challenges 
when the target space and the base space are compactified, 
so that winding modes appear.

When the base space is a torus, for example, 
there are modular transformations on the torus 
which do not change the geometry of the torus, 
and are thus expected to be symmetries of the theory.
However, in general the partition function of a chiral boson 
cannot be modular invariant.
One has to accept partition functions 
which change only by an overall factor under 
modular transformations, 
like $\vth$-functions.
(After all, the partition function is essentially a wave function.)

A geometrical interpretation of this modular non-invariance of 
the chiral boson theory is that 
the quantum theory of a chiral boson depends on the choice of 
geometrical structures other than those uniquely determined by
the Riemannian metric and complex structure.
(Here the relevant geometrical structure is the spin structure.)
Isometry is hence not necessarily a symmetry of the theory.
(See more comments on this in the last section.)
Depending on the applications in mind 
(the physical properties of the system of interest),
there may be certain restrictions on the spin structures. 
In this work, 
we will not be restricted to particular applications 
and consider the full generality.

\section{Holomorphic Decomposition}
\label{Holo-decomp}

In this paper, 
we consider a chiral boson living on a flat 2-torus $\Sigma$ defined by
the equivalence relations
\be
z \sim z + m + n\tau \qquad (m, n \in \mathbb{Z})
\ee
in terms of the complex coordinate
\be
z = \sigma_1 + i \sigma_2
\label{zs}
\ee
for a given modular parameter $\tau = \tau_1 + i\tau_2$
($\tau_2 > 0$).
The self-duality condition is $\del_{\bar{z}}\phi(z) = 0$.

The target space of the scalar field $\phi$ is a circle.
We normalize $\phi$ so that it is defined up to multiples of $2\pi$,
\be
\phi \sim \phi + 2\pi.
\label{phi-equiv}
\ee

Roughly speaking, as the non-chiral boson can be viewed 
as the combination of a chiral boson plus an anti-chiral boson, 
the partition function of a non-chiral boson can be decomposed 
into the product of the contribution of the chiral boson
and that of the anti-chiral boson.
We refer to it as a holomorphic decomposition.

For a non-chiral boson with the free-field action 
\be
\label{HNSa2}S_0[\phi] = \frac{1}{\pi g^2} \int dzd\zb \; \del_z\phi \del_{\zb}\phi,
\ee
to selectively turn on the chiral and anti-chiral parts of the scalar,
we introduce two sources $A_{\zb}$ and $A_z$ via interaction terms as
\footnote{
In our convention, $\int dz d\bar z=2\tau_2$.
}
\be
\label{HNSa1}S_{A}[\phi, A] = 
\frac{1}{\pi g}\int dzd\bar{z} \; (A_{\bar{z}} \del_z\phi + A_z \del_{\bar{z}}\phi).
\ee
The coupling constant $g$ is the only free parameter of the theory.
It can be normalized to $1$ by scaling $\phi$,
so we can also interpret it as the inverse radius of the target-space circle.

Naively, 
the partition function
\be
Z_{\mbox{\small non-chiral}}[A_{\bar{z}}, A_z] 
= \int D\phi \; e^{-(S_0[\phi] + S_A[\phi, A])}
\ee
is expected to be of the form of a product
\be
Z_{\mbox{\small non-chiral}}[A_{\bar{z}}, A_z] 
= {Z}_{\mbox{\small hol}}[A_{\bar{z}}]{Z}_{\mbox{\small anti-hol}}[A_{z}]
\label{naive-hol-decomp}
\ee
of a holomorphic functional of $A_{\bar{z}}$ and
an anti-holomorphic functional of $A_{z}$.
The partition function of the chiral boson is expected to be 
given by the holomorphic functional $Z_{\mbox{\small hol}}[A_{\bar{z}}]$.

A serious problem is that,
when the base space has nontrivial topology,
the naive holomorphic decomposition (\ref{naive-hol-decomp})
is replaced by a more complicated decomposition of the form
\be
Z_{\mbox{\small non-chiral}}[A_{\bar{z}}, A_z] = 
\sum_s {Z}^{(s)}_{\mbox{\small hol}}[A_{\bar{z}}]{Z}^{(s)}_{\mbox{\small anti-hol}}[A_z].
\label{hol-decomp}
\ee
The index $s$ in (\ref{hol-decomp})
is the label for the spin structure on the torus $\Sigma$.
This means that there is an ambiguity in defining the partition function 
of a chiral boson:
it can be any of the holomorphic factors $Z^{(s)}_{\mbox{\small hol}}$.

In fact, we will see below that (\ref{hol-decomp}) is not exactly correct.
There exists an additional overall factor on the right hand side
that depends on both $A_{\bar{z}}$ and $A_z$.
Nevertheless this intuition leads us to a way to define the partition function 
of a chiral boson from that of a non-chiral boson.
The calculation below is a generalization of 
the work of Ref. \cite{Henningson:1999dm}.

It is natural to decompose the source $A$
\be
A = A^{(0)} + \tilde{A},
\ee
into the zero mode $A^{(0)}$ and oscillator modes $\tilde{A}$ 
since they decouple in the quadratic Lagrangian.
The partition function 
\be
Z[A] = Z_0[A^{(0)}] \tilde{Z}[\tilde{A}]
\ee
is a product of 
a function $Z_0[A^{(0)}]$ of the zero mode of $A$
and a functional $\tilde{Z}[\tilde{A}]$ of the non-zero modes of $A$.
The zero-mode part of the partition function $Z_0[A^{(0)}]$
is a sum over winding modes of the scalar field $\phi$, 
while the non-zero-mode part $\tilde{Z}[\tilde{A}]$ comes from 
an integral over all non-zero modes of $\phi$.
Special attention is paid to the factor $Z_0[A^{(0)}]$ depending on the zero mode $A^{(0)}$,
as it is where the nontrivial dependence on the spin structure resides.

\subsection{$Z_0[A^{\z}]$}

Due to the equivalence relation (\ref{phi-equiv}),
the periodic boundary conditions for $\phi$ over the two cycles of the torus $\Sigma$ are
\be
\label{peri}
\phi(z+m+n\tau, \zb+m+n\bar{\tau}) = \phi(z, \zb) + 2\pi(m\om_1 + n\om_2)
\ee
for arbitrary winding numbers $\om_1, \om_2 \in \mathbb{Z}$.
The winding mode of the chiral boson is thus given by
\be
\label{zphi}
\phi^{\z} =
\frac{\pi i}{\tau_2}\left[\om_1(\bar \tau z - \tau\bar z)
-\om_2(z-\bar z)\right].
\ee

The zero-mode part of the partition function $Z_0[A^{\z}]$ 
is computed by substituting eq.\eqref{zphi} into the action $S_0 + S_A$
and summing over all winding numbers $\om_1, \om_2$.
It is
\be
Z_0[A^{\z}] =
\sum_{\om_1, \om_2 \in \mathbb{Z}}
\exp\Big[-\frac{2\pi}{ g^2 \tau_2}|\om_2-\om_1\tau|^2
+\frac{2i}{g} A^{(0)}_{z}(\om_1\tau-\om_2) 
-\frac{2i}{g} A^{(0)}_{\zb}(\om_1\bar\tau-\om_2) \Big].
\ee
Applying the Poisson resummation formula 
\be
\sum_{m=-\infty}^{\infty}\exp\left[-\frac{\pi(m-b)^2}{a}\right]
=a^{1/2}\sum_{n=-\infty}^{\infty}\exp(-\pi a n^2+2\pi i bn)
\label{Poisson}
\ee
to the dummy variable $\om_2$,
we find
\be
\label{ww}
Z_0[A^{\z}]
=
\mathcal W[A^{\z}]
\sum_{\om_1, \om_2 \in\mathbb{Z}}
\hat{h}\left(\frac{g\om_2}{2}+\frac{\om_1}{g},y\bigg|\tau\right)
\overline{\hat{h}\left(\frac{g\om_2}{2}-\frac{\om_1}{g}, -y\bigg|\tau\right)},
\ee
where
\bea
\mathcal W[A^{\z}]
&\equiv&
\sqrt{\frac{g^2\tau_2}{2}} \;
\exp\left[\frac{\pi}{2\tau_2}(y+\bar{y})^2\right],\\
\hat{h}(u,z|\tau)&\equiv&\exp(i \pi \tau u^2 + i 2\pi uz),\\
y&\equiv& \frac{i\tau_2}{\pi} A^{\z}_{\zb}.
\eea
Here we assume that $\bar{y} = -\frac{i\tau_2}{\pi} A^{\z}_{z}$.

For special values of the coupling
\footnote{
Incidentally, there is a symmetry of the partition function of 2-dimensional scalar,
\be
g \rightarrow \frac{2}{g},
\nn
\ee
with a simultaneous change of sign for $A^{\z}_z$.
}
\be
g = \sqrt{k_1k_2} \qquad
\mbox{or} \qquad
g = \frac{2}{\sqrt{k_1k_2}},
\ee
where $k_1, k_2$ are positive integers and $k_2$ is even,
the partition function can be written as a finite sum over 
products of holomorphic functions of $A^{\z}_{\bar{z}}$ 
and anti-holomorphic functions of $A^{\z}_{z}$,
up to the anomalous factor ${\cal W}[A^{\z}]$.
Using the following identity
\be
\sum_{n_2=0}^{k_2-1} e^{2\pi i(m_1-m_2)n_2 / k_2}= 
k_2 \sum_{m'=-\infty}^{\infty}\delta^{m'k_2}_{m_1-m_2},
\ee
we simplify the expression above for the partition function as
\beqa
\label{theW1}
Z_0[A^{\z}] &=&
k_2^{-1}\mathcal W[A^{\z}]
\sum_{n_1 = 0}^{k_1-1}\sum_{n_2 = 0}^{k_2-1} 
\vth\left[\frac{n_1}{k_1}\atop{\frac{n_2}{k_2}}\right]
\left(\sqrt{\frac{k_1}{k_2}}y; \frac{k_1}{k_2}\tau\right) 
\overline{\vth\left[\frac{n_1}{k_1}\atop{\frac{n_2}{k_2}}\right]
\left(\pm\sqrt{\frac{k_1}{k_2}}y; \frac{k_1}{k_2}\tau\right)}.
\eeqa
For the anti-holomorphic function,
the sign $\pm$ in front of $y$ should be
$+$ for $g = \sqrt{k_1k_2}$ and $-$ for $g = \frac{2}{\sqrt{k_1k_2}}$, respectively. 
This is a generalization of the result in \cite{AlvarezGaume:1987vm,Henningson:1999dm}.

The partition function of a chiral boson is then identified 
with the holomorphic factor in this decomposition, 
which is a $\vth$ function
\be
Z^{\mbox{\small chiral}}_0[A^{\z}] = 
\vth\left[\frac{n_1}{k_1}\atop{\frac{n_2}{k_2}}\right]
\left(\sqrt{\frac{k_1}{k_2}}y; \frac{k_1}{k_2}\tau\right).
\label{chiralZ0}
\ee
Note that here the modular parameter appearing in the $\vth$ function 
is not necessarily the same as the modular parameter of the spacetime,
as we often see in the literature, 
but can differ from it by a fractional factor $k_1/k_2$.
\footnote{
In Ref.\cite{Henningson:1999dm},
the partition function for a chiral boson is restricted to be 
the cases for $k_1 = k_2 = 2$ so that it is
$\vth\left[n_1/2\atop n_2/2\right]\left(y; \tau\right)$
for $n_1, n_2 = 0, 1$.
}
This is a possible generalization of the partition function for chiral boson 
that was not emphasized in the past. 
We will see that the same result is obtained 
in the other two approaches for computing partition functions for a chiral boson 
that we will discuss below.
In fact, as the partition function is a function of the source field $A_{\bar{z}}$,
the modular parameter appearing in the $\vth$ function is supposed to be 
the parameter characterizing the space of $A^{\z}_{\bar{z}}$, 
rather than the space of $z$.
Upon a modular transformation on the space of $z$,
the spacetime is exactly the same 2-torus as before the transformation, 
but in general 
the partition function is not invariant.
That is, there are anomalies in the modular transformations
for the quantum theory of a chiral boson.


The holomorphic decomposition of the partition function for a non-chiral boson is not unique.
For any choice of the parameters $k_1, k_2$, 
there is a holomorphic decomposition.
Furthermore, 
the Poisson resummation formula can be applied to the other winding number $\om_1$
instead of $\om_2$.
Following similar calculations as above,
we find, 
for positive integers $k_1, k_2$ ($k_2$ must be even),
\beqa
Z_0[A^{\z}]
&=&
k_2^{-1} \mathcal W'[A^{\z}]
\sum_{n_1 = 0}^{k_1-1}\sum_{n_2 = 0}^{k_2-1} 
\vth\left[\frac{n_1}{k_1}\atop{\frac{n_2}{k_2}}\right]
\left(\sqrt{\frac{k_1}{k_2}}\frac{y}{\tau}; -\frac{k_1}{k_2}\frac{1}{\tau}\right) 
\overline{\vth\left[\frac{n_1}{k_1}\atop{\frac{n_2}{k_2}}\right]
\left(\pm\sqrt{\frac{k_1}{k_2}}\frac{y}{\tau}; -\frac{k_1}{k_2}\frac{1}{\tau}\right)},
\nn
\\
\eea
where the anomalous factor is
\be
\mathcal{W}'[A^{\z}] = 
\sqrt{\frac{g^2\tau_2}{2|\tau|^2}} \;
\exp\left[\frac{\pi |\tau|^2}{2\tau_2} (y/\tau + \bar y/\bar \tau)^2\right].
\ee

Using the modular transformation property (\ref{mod-transf-th}) of the $\vth$ function,
we find
\beqa
\label{theW2}
Z_0[A^{\z}]
&=&
k_1^{-1} \mathcal W[A^{\z}]
\sum_{n_1 = 0}^{k_1-1}\sum_{n_2 = 0}^{k_2-1} 
\vth\left[\frac{n_2}{k_2}\atop{\frac{n_1}{k_1}}\right]
\left(\sqrt{\frac{k_2}{k_1}}y; \frac{k_2}{k_1}\tau\right) 
\overline{\vth\left[\frac{n_2}{k_2}\atop{\frac{n_1}{k_1}}\right]
\left(\pm\sqrt{\frac{k_2}{k_1}}y; \frac{k_2}{k_1}\tau\right)}.
\eea
The anomalous factor is the same as that in (\ref{theW1}).
Despite the difference in explicit expressions,
both (\ref{theW1}) and (\ref{theW2}) tell us that 
the partition function of a chiral boson is the $\vth$ function 
with arbitrary rational characteristics.
The chiral boson's partition function read off from (\ref{theW2}) 
is equivalent to (\ref{chiralZ0}) but with the indices 1 and 2 interchanged.

This approach of holomorphic decomposition suffers several disadvantages.
First, the Lagrangian of a non-chiral boson is assumed.
For a given Lagrangian for an interacting chiral boson theory, 
e.g. that of Ref. \cite{Pasti:1997gx},
it is not clear how to define the Lagrangian of a non-chiral boson 
whose partition function can be holomorphically decomposed to 
define the partition function of the chiral boson.
Secondly, even if the non-chiral theory is known, 
the holomorphic decomposition involves an ``anomalous'' factor ${\cal W}$ \cite{Henningson:1999dm}.
It is not clear how to determine this factor 
and as a result the holomorphic decomposition is strictly speaking ambiguous.
This problem cannot be avoided by defining the holomorphic decomposition 
in terms of the holomorphicity of the modular parameter $\tau$.
To define the holomorphicity of $\tau$, 
we have to decide first what is independent of $\tau$.

Note that $k_1 = k_2 = 2$ gives $g^2 = 1$ and $g^2 = 4$.
The literature has mostly focused on the case 
$\vth\col{n_1/2}{n_2/2}$ with $n_1, n_2 = 0, 1$.
The case $g^2 = 2$ is special in that the theory is self-dual,
and the only possibility is $k_1 = 1, k_2 = 2$.
In general, there are many pairs of $(k_1, k_2)$ corresponding to the same value of $g^2$.
(For example, $g^2 = 4$ can be given by $(k_1 = 2, k_2 = 2)$ and $(k_1 = 1, k_2 = 4)$.)

So far we have found the use of $\vth$ functions $\vth\col{\a}{\b}$
as the partition function of a chiral boson
only for $\a, \b$ being rational numbers.
In fact it is easy to generalize them to arbitrary real numbers 
by introducing a twisting of the boundary condition of $\phi$.
Equivalently, we can introduce a constant connection 1-form $C$ so that
$d\phi$ is replaced by $d\phi + C$ in the Lagrangian.
Following the same calculation above,
the two components $C_z$ and $C_{\bar{z}}$ will show up in 
the characteristics of the $\vth$ function.

If we think of a chiral boson as the bosonization of
a chiral fermion with boundary conditions 
defined by the periodic or anti-periodic boundary conditions,
the relevant $\vth$ function would only have parameters $(\a, \b)$ 
for $\a, \b = 0, 1/2$.
On the other hand, 
if the chiral fermion is charged and admits twisted boundary conditions,
the partition function is allowed to be a $\vth$ function with irrational parameters.

\subsection{$\tilde{Z}[\tilde{A}]$}

To compute the non-zero-mode part of partition function, 
we pick a basis of 0-forms 
\be
B = \left\{ b_{mn} \equiv \exp\left(\frac{\pi[m(\zb\tau-z\bar\tau)+n(z-\zb)]}{\tau_2}\right) \right\}
= B^{\z} \cup \tilde B.
\label{bmn}
\ee
For the non-zero modes in $\tilde{B}$, 
the index $(mn)$ does not take the value $(00)$.
Here we focus on the non-zero modes $\tilde B$.
The scalar $\phi$ can be decomposed in this basis
\be
\phi = \phi^{\z} + \tilde \phi, \qquad
\tilde \phi = \sum_{m, n\in\mathbb{Z}} \Phi_{mn}b_{mn}
\ee
so that 
\beqa
\partial_z \tilde \phi = i \sum_{m, n \in \mathbb{Z}} \Phi_{mn}b_{mn}\Pi_z^{mn}, 
\qquad
\partial_{\zb} \tilde \phi = i \sum_{m, n \in \mathbb{Z}} \Phi_{mn}b_{mn}\Pi_{\zb}^{mn},
\eeqa
where
\beqa
\Pi_z^{mn}=-\frac{\pi(m\bar\tau-n)}{i\tau_2},
\qquad
\Pi_{\zb}^{mn}=\frac{\pi(m\tau-n)}{i\tau_2}.
\eeqa
Similarly, we can expand $\tilde A$ as
\beqa
\tilde A_z = \sum_{m, n \in \mathbb{Z}} \tilde A_z^{mn}b_{mn},
\qquad
\tilde A_{\zb} = \sum_{m, n \in \mathbb{Z}} \tilde A_{\zb}^{mn}b_{mn}.
\eeqa
Substituting $\tilde{A}$ and $\partial \tilde{\phi}$ into $\tilde{Z}[\tilde{A}]$ 
and integrating over $\Phi$ gives
\beqa
\notag \tilde{Z}[\tilde{A}]&=&
\left[\det\left(\frac{4\tau_2\Pi_z\Pi_{\zb}}{\pi g^2}\right)\right]^{-1/2}
\tilde{\cal W}[\tilde{A}]
\exp\left(\sum_{mn}\frac{m\tau_2\tilde A_{\zb}^{mn}\tilde A_{\zb}^{-m-n}}
{\Pi_{\bar z}^{mn}}
\right)
\overline{\exp\left(\sum_{mn}\frac{m\tau_2\tilde A_{\zb}^{mn}\tilde A_{\zb}^{-m-n}}
{\Pi_{\bar z}^{mn}}
\right)}\\
&=&
\frac{\tilde{\cal W}[\tilde{A}]}{\sqrt{\pi\tau_2}g}
\left[\frac{1}{\eta}
\exp\left(\sum_{mn}\frac{m\tau_2\tilde A_{\zb}^{mn}\tilde A_{\zb}^{-m-n}}
{\Pi_{\bar z}^{mn}}
\right)\right]
\overline{\left[\frac{1}{\eta}
\exp\left(\sum_{mn}\frac{m\tau_2\tilde A_{\zb}^{mn}\tilde A_{\zb}^{-m-n}}
{\Pi_{\bar z}^{mn}}
\right)\right]},
\label{KK-Z-HNS}
\eeqa
where 
\be
\tilde{\cal W}[\tilde{A}] =
\exp\left(-\frac{\tau_2}{2\pi}\sum_{mn}
(\tilde A_z-\tilde A_{\zb})^{mn}(\tilde A_z-\tilde A_{\zb})^{-m-n}\right),
\ee
and
we have applied the zeta function regularization to compute the determinant in the last equality
\beqa
\left[\det\left(\frac{4\tau_2\Pi_z\Pi_{\zb}}{\pi g^2}\right)\right]^{-1/2}
=\frac{1}{\sqrt{\pi\tau_2}g \eta\bar\eta}.
\eeqa
Here $\eta$ is the Dedekind eta function.

The non-zero-mode part of the partition function for a chiral boson on 2-torus 
can be read off as the holomorphic part of (\ref{KK-Z-HNS}) to be
\be
\tilde{Z}_{\mbox{\small chiral}}[\tilde{A}] = 
\frac{1}{\eta}
\exp\left(\sum_{m,n}\frac{m\tau_2\tilde A_{\zb}^{mn}\tilde A_{\zb}^{-m-n}}
{\Pi_{\bar z}^{mn}}
\right).
\label{ZKK-HNS}
\ee
It is independent of the spin structure.

\section{Holomorphic Line Bundle}
\label{Holo-bundle}

In this section we review the approach of \cite{Witten:1996hc,Belov:2006jd}
to compute the partition function of a chiral boson.
Consider the manifestly Lorentz-invariant action of a non-chiral boson $\phi$ 
coupled to an external vector field $A$ in the form of a $U(1)$ connection
\be
S[\phi, A] = 
\frac{1}{\pi g^2}\int_{\Sigma} dz d\bar{z} 
\left[
(\del_z \phi + g A_z/2)(\del_{\bar{z}} \phi + g A_{\bar{z}}/2)
- g\phi F_{z\bar{z}}/2
\right],
\label{L1}
\ee
where the last term is added to decouple the anti-chiral component $\del_{\zb}\phi$.
After integration by parts, the action is
\be
S[\phi, A] = \frac{1}{\pi g^2} \int_{\Sigma} dz d\bar{z}
\left(\del_z\phi\del_{\bar{z}}\phi + g A_{\zb} \del_{z}\phi + g^2 A_z A_{\bar{z}}/4 \right).
\label{L2}
\ee
This action first appeared in \cite{Jackiw:1984zi}
as the action for the bosonization of a chiral fermion. 

The transformations
\be
\d \phi = - g\lam/2, \qquad
\d A = d\lam
\label{gauge-transf}
\ee
for a generic function $\lam$
are not symmetries because of the last term of the action (\ref{L1}).
This does not lead to pathologies of the quantum theory
because $A$ is an external field without dynamics.

The partition function 
\be
Z[A] = \int D\phi e^{-S[\phi, A]}
\ee
is not independent of $A_{z}$
even though the coupling of $A_{z}$ to $\del_{\zb}\phi$ is cancelled, 
due to the last term of eq.(\ref{L2}).
Similar to the previous section,
the partition function 
can be factorized as
\be
Z[A_{{\zb}}] = Z_0[A^{(0)}_{{\zb}}]\tilde{Z}[\tilde{A}_{{\zb}}].
\ee

Consider the full space of configurations of $A$ 
as the base space of a line bundle on which $Z[A]$ is a section.
Define the covariant derivatives for the line bundle as
\bea
\frac{D}{D A_z} 
= \frac{\delta}{\delta A_z} +\frac{ A_{\bar{z}}}{4\pi},
\qquad
\frac{D}{D A_{\bar{z}}} 
= \frac{\delta}{\delta A_{\bar{z}}} - \frac{ A_z}{4\pi}.
\label{covder-2}
\eea
For the action (\ref{L2}),
$Z[A]$ satisfies two differential equations
\bea
&\frac{D}{D A_{z}} Z[A] = 0, \label{DAzZ} \\
&\left[
\del_z\frac{D}{D A_z} + \del_{\bar{z}}\frac{D}{D A_{\bar{z}}} -\frac{ F_{z\bar{z}}}{2\pi}
\right] Z[A] = 0. \label{DAzbarZ}
\eea
We will compute the partition function $Z[A]$
by solving these differential equations,
with suitable boundary conditions.

The two differential equations (\ref{DAzZ}), (\ref{DAzbarZ}) 
are sufficient to determine how $Z[A]$ 
transforms under the transformation
\bea
\delta A_{\zb} &=& -\del_{\zb} \eps_{z}, 
\label{da} \\
\delta A_{z} &=& \lam,
\eea
where $\eps_z$ and $\lam$ are independent functions.
One can transform an arbitrary configuration of $A$ to
\be
A_{\zb} = A^{(0)}_{\zb} 
\qquad \mbox{and} \qquad
A_{ z} = 0,
\ee
where $A^{(0)}_{\zb}$ represents the zero mode of $A_{\zb}$.
(In fact, there is no reason to consider $A_z \neq 0$ 
while we are not coupling it to the anti-chiral field $\del_{\zb}\phi$.)
That is, if we can determine $Z[A_{\zb}]$ on the space of $A^{(0)}_{\zb}$,
we can uniquely determine $Z[A]$ as a solution to (\ref{DAzZ}) and (\ref{DAzbarZ}).

Since the change (\ref{da}) of $A_{\zb}$ is 
the same as that due to a gauge transformation (\ref{gauge-transf}),
we realize that there are 
large gauge transformations changing $A^{(0)}_{\zb}$
by a vector on a lattice $L$, 
\be
A^{(0)}_{\zb} \rightarrow A^{(0)}_{\zb} + \frac{2\pi i}{g\tau_2 }(m\tau-n)
\qquad 
(m, n \in \mathbb{Z}),
\label{equiv-A0}
\ee
so $A^{(0)}_{\zb}$ can be viewed as a complex coordinate on
the quotient $\mathbb{C}/L$,
which 
will be denoted as $J_{\Sigma}$.
It is sufficient to determine $Z[A_{\zb}]$ on 
$J_{\Sigma}$
in order to uniquely determine $Z[A]$.

It is impossible for $Z[A^{(0)}_{\zb}]$ to be a well-defined function on $J_{\Sigma}$.
However, it is still possible to demand that
a line bundle equipped with the covariant derivatives 
(\ref{covder-2})
be reduced to a holomorphic line bundle on $J_{\Sigma}$
of which $Z[A^{(0)}_{\zb}]$ is a section.
This is acceptable because the partition function is defined up to a phase.
In fact, there are more than one holomorphic line bundles
equipped with the same covariant derivatives 
(\ref{covder-2}).

Before getting into the details of actual calculation,
we comment that this approach, like the previous approach,
is not applicable to more general theories of chiral bosons.
In this approach, 
we assume that the chiral boson is always defined as the chiral part of a non-chiral boson.
However, 
it is not clear how to apply this approach to a generic action for a non-chiral boson.
\footnote{
This approach was generalized and applied to the chiral WZW model \cite{Witten:1991mm}.
}

The gauge transformation of the source field $A$ plays a crucial role in this approach,
so we are led to replace derivatives $d \phi$ by the covariant derivatives $d \phi + A$
everywhere in the action.
For a generic action,
this does not necessarily yield the desired coupling between 
the chiral boson and the source field $A$
so that $A_{z}$ is decoupled from the anti-chiral component of $\phi$
(which may be a deformed version of $\del_{\zb}\phi$).

More specifically, the success of this approach relies on the following properties.
First, 
one has to know the action of the non-chiral boson corresponding to
the chiral boson of interest.
Second, the action should equal the scalar field action 
plus a source term that couples only to the chiral field,
up to an irrelevant term that vanishes 
when we set the irrelevant source field $A_{z}$ to zero.
Third,
there exists a transformation of the fields such that
the induced change of the action only depends on the source $A$ but not the scalar $\phi$,
so that the partition function can be interpreted as a section of a line bundle over 
the space of $A$, 
with the equivalence relations analogous to (\ref{equiv-A0}) 
defined by these transformations.

\subsection{$Z_0[A^{\z}]$}

We would like to give an explicit expression of the partition function 
of the chiral boson on $\Sigma$.
To proceed,
we decompose the source $A_{\zb}$ into 
the zero mode (constant) and non-zero modes
\be
A_{\zb} = A^{(0)}_{\zb} + \tilde{A}_{\zb}.
\ee
There is no linear term in $z, \bar{z}$ because
they would imply a zero mode in the field strength $F$,
leading to a nontrivial twist $\int_{\Sigma} F$ of the $\phi$-bundle over $\Sigma$.
Instead, $A$ is always a connection for a trivial bundle
because its field strength $F = dA = d\Lambda$
is the same as the field strength of the connection
$\Lambda \equiv d\phi + A$,
which is gauge invariant and globally well defined.

The subtlety of the calculation of partition function 
resides in choosing a holomorphic line bundle ${\cal L}^{(s)}$ on $J_{\Sigma}$.
The connection on the holomorphic line bundle ${\cal L}^{(s)}$ 
is independent of $s$ and given by 
eq.(\ref{covder-2}).
The curvature is
\be
\left[ \frac{D}{DA_z(z, \bar{z})}, \frac{D}{DA_{\bar{z}}(z', \bar{z}')} \right]
= -\d^2(z - z',\zb-\zb') \frac{1}{2\pi}.
\label{curvature1}
\ee
Restricting the bundle to $J_{\Sigma}$, 
it is useful to define the zero modes of the covariant derivatives as
\bea
\frac{D}{DA_z^{(0)}} \equiv \left.\int_{\Sigma} dz d\bar z
\frac{D}{D A_z(\s)}\right|_{\mathrm{zero~mode}} = 
\frac{\del}{\del A_z^{(0)}} + \tau_2 \frac{ A_{\bar{z}}^{(0)}}{2\pi}, 
\label{DDAz} \\
\frac{D}{DA_{\bar{z}}^{(0)}} \equiv \left.\int_{\Sigma} dz d\bar z
\frac{D}{D A_{\bar{z}}(\s)} \right|_{\mathrm{zero~mode}}= 
\frac{\del}{\del A_{\bar{z}}^{(0)}} - \tau_2 \frac{ A_{z}^{(0)}}{2\pi}.
\label{DDAzbar}
\eea
Eq. (\ref{DAzZ}) thus implies
\be
\left(\frac{\del}{\del A_{ z}^{(0)}} + \tau_2 \frac{A_{{\zb}}^{(0)}}{2\pi}\right) Z_0[A^{(0)}] = 0.
\label{DAzZ2}
\ee
The other differential equation (\ref{DAzbarZ}) is trivial for $Z_0[A^{\z}]$.

For convenience,
define a new complex coordinate on $J_{\Sigma}$
\be
y = \frac{i \tau_2 A^{(0)}_{\zb}}{\pi}, 
\qquad
\bar y= - \frac{i \tau_2 A^{(0)}_{z}}{\pi}.
\label{y-in-A}
\ee
The equivalence relations are 
\be
y\sim y+\frac{2}{g}(m+n\tau) \qquad (m, n \in \mathbb{Z}).
\ee

The covariant derivatives 
(\ref{DDAz}), (\ref{DDAzbar}) are now expressed as
\be
D_{\bar y} = \frac{\del}{\del \bar y} - \frac{F}{2} y, \qquad
D_{y} = \frac{\del}{\del y} + \frac{F}{2} \bar y, 
\label{DyDy}
\ee
which define a line bundle over $J_{\Sigma}$
with the field strength
\be
F = -\frac{\pi}{\tau_2}.
\ee
The twisted periodic boundary conditions over an $(m, n)$-cycle of 
$J_{\Sigma}$
are given by a phase factor $U_{(m, n)}$ as
\be
U_{(m, n)}D_i(y; \bar y) U^{-1}_{(m, n)} = 
D_i\left(y+ \frac{2}{g}(m+ n\tau); \bar y+ \frac{2}{g}(m+ n\bar{\tau})\right).
\ee
From (\ref{DyDy}), we find 
\be
U_{(m, n)} = e^{-\frac{F}{g}(m+n\bar \tau)y+\frac{F}{g}(m+n\tau)\bar y+f(m, n)},
\label{U1}
\ee
where $f(m, n)$ is an arbitrary function of $m, n$
(and of $\tau$, but independent of $y$ and $\bar{y}$).

As a fundamental representation,
a section $Y$ of the bundle 
should satisfy the twisted periodic boundary condition
\be
U_{(m, n)}Y(y, \bar{y}) = Y\left(y+\frac{2}{g}(m+n\tau), \bar{y}+\frac{2}{g}(m+n\bar\tau)\right).
\ee
Moreover, $Y$ should respect $D_{\bar{y}} Y=0$, which implies 
\be
Y(y, \bar y) = e^{\frac{F}{2}y\bar y-\frac{F}{2}y^2} Y_0(y),
\ee
then the twisted periodic boundary condition on $Y_0$ is
\be
e^{- i4\pi ny/g - i4\pi n(m+n\tau)/g^2 + f(m, n)}Y_0(y) 
= Y_0\left(y+\frac{2}{g}(m+n\tau)\right).
\label{twist-Y}
\ee
In general, choosing a specific $f(m, n)$ corresponds to 
choosing a holomorphic line bundle as well as a specific spin structure 
(cf. eq.\eqref{thetaZ}) which gives
\be
Y_0(y) = \vth\col{\a}{\b}(ay; b\tau),
\ee
where $a, b$ satisfy the relations 
\be
a^2 = b, \qquad
\frac{2a}{g} \in \mathbb{Z}, \qquad
\frac{2a}{bg} \in \mathbb{Z}.
\ee
Solutions to these relations can be parametrized by two integers $k_1, k_2$:
\be
a = \sqrt{\frac{k_1}{k_2}}, \qquad
b = \frac{k_1}{k_2}, \qquad
g = \frac{2}{\sqrt{k_1 k_2}}.
\ee
This is in agreement with the result of holomorphic decomposition in eq.(\ref{chiralZ0}).

For the special case $g = 2$,
the total flux over $J_{\Sigma}$ is a unit quantum
consistent with treating $J_{\Sigma}$ as a 2-torus.
Holomorphic line bundles with unit flux on 2-torus are fully understood.
The moduli space is itself also a 2-torus.
It is also known that each holomorphic line bundle ${\cal L}^{(s)}$
has a unique holomorphic section, 
which in a certain trivialization 
is given by the $\vth$ function $\vth\col{\a}{\b}(y|\tau)$,
and the parameters $\a, \b \in \mathbb{R}/\mathbb{Z}$ 
label the choice of the holomorphic line bundle.

In general,
the index of the line bundle can be expressed as $s = (\a, \b)$, 
and we have
\be
Z_0^{(s)}[A^{(0)}_{\bar{z}}] = {\cal N}^{(s)} 
e^{\frac{F}{2}y\bar y-\frac{F}{2}y^2 }
\vth\col{\a}{\b}\left(\sqrt{\frac{k_1}{k_2}}y; \frac{k_1}{k_2}\tau\right)
\ee
for some normalization factor ${\cal N}^{(s)}$.
In this approach the characteristics $(\a, \b)$ of the $\vth$ function 
are not restricted to be rational,
unless further assumptions are made.
Only the cases of $\a, \b = 0, 1/2$ were considered in \cite{Witten:1996hc},
corresponding to having either the periodic or antiperiodic boundary conditions 
for the spin structure on 2-torus.
It was later generalized in \cite{Belov:2006jd} to more general holomorphic line bundles.

\subsection{$\tilde{Z}[\tilde{A}]$}

Next we find the partition function 
$\tilde{Z}[\tilde{A}_{\bar{z}}]$
for the non-zero modes
by solving the differential equations (\ref{DAzZ}), (\ref{DAzbarZ}).
Consider the ansatz
\be
\tilde{Z}[\tilde{A}_{\bar{z}}] = 
e^{-\frac{1}{4\pi}\int dzd\zb \tilde{A}_{z}(z, \zb) \tilde{A}_{\bar{z}}(z, \zb)}
e^{\int dzd\zb \int dz' d\zb' \; K(z-z', \zb-\zb') \tilde{A}_{\zb}(z, \zb) \tilde{A}_{\zb}(z', \zb')},
\label{ansatz-KKZ}
\ee
where $K$ is allowed to depend on $\tau$ and $A^{(0)}_{\zb}$.
While (\ref{DAzZ}) is already solved by the ansatz above,
eq.(\ref{DAzbarZ}) implies that
\be
\int dz' d\zb' \; \left(
\del_{\zb} K(z-z', \zb-\zb')\tilde{A}_{\zb}(z', \zb') \right) 
- \frac{1}{4\pi} \del_z \tilde{A}_{\zb}(z, \zb) 
= 0.
\ee
This is solved if
\be
\del_{\zb}K(z-z', \zb-\zb') = - \frac{1}{4\pi} \del_z \d(z-z', \zb-\zb').
\label{Diff-K}
\ee

Given the ordinary Green's function $G$
for the Laplace operator $\del_z\del_{\zb}$ on $\Sigma$,
\be
\del_z\del_{\zb} G(z-z', \zb-\zb') = \d(z-z', \zb-\zb') + \mbox{constant},
\label{Laplace}
\ee
where the constant on the right hand side is needed on compact base space,
then $K$ is immediately solved by
\be
K(z-z', \zb-\zb') = - \frac{1}{4\pi} \del_z^2 G(z-z', \zb-\zb')
+ K_0,
\label{KG}
\ee
where $K_0$ is a holomorphic function.
Note that the only well-defined holomorphic function on $\Sigma$ is constant,
and constant $K_0$ makes no contribution to the ansatz (\ref{ansatz-KKZ})
because the non-zero modes $\tilde{A}_{\bar{z}}$ integrate to zero.
Hence we can set $K_0$ to zero without loss of generality.


\section{Path Integral for FJ Model}
\label{Dir-calc}

In this section we try to reproduce the results of the previous sections
on the partition function for a chiral boson, 
by calculating the path integral of the Floreanini-Jackiw (FJ) action \cite{Floreanini:1987as},
instead of calculating the path integral for a non-chiral boson.
The FJ action has Lorentz symmetry although it is not manifestly Lorentz-invariant.
It can be derived from the Lorentz-invariant action of Siegel \cite{Siegel:1983es}
by gauge fixing.
On the other hand, the Lorentz symmetry is broken by 
the geometry of the base space.

The FJ action has to be modified in order to reproduce 
the same partition function as the other two approaches discussed above.
(The original FJ action produces a sum over different $\vth$-functions,
instead of a single $\vth$-function,
as its partition function.)
Before proposing our modification to the FJ action, 
let us introduce some notations.
On the torus it is natural and convenient to define 
the basis of one-forms dual to the two cycles 
(called $\Ac$-cycle ${\cal C}_\Ac$ and $\Bc$-cycle ${\cal C}_\Bc$) as
\be
E_\Ac = d\sigma_\Ac,
\qquad
E_\Bc = d\sigma_\Bc,
\ee
where
\be
\sigma_\Ac = \frac{-\bar{\tau} z + \tau \bar{z}}{\tau - \bar{\tau}},
\qquad
\sigma_\Bc = \frac{z - \bar{z}}{\tau - \bar{\tau}},
\ee
so that
\be
\oint_{{\cal C}_\Ac} E_\Ac = 1 = \oint_{{\cal C}_\Bc} E_\Bc, 
\qquad
\oint_{{\cal C}_\Ac} E_\Bc = 0 = \oint_{{\cal C}_\Bc} E_\Ac,
\ee
and
\be
\int E_\Ac\wedge E_\Bc = 1.
\ee
The exterior derivative is then
\be
d = E_\Ac \del_\Ac + E_\Bc \del_\Bc,
\ee
where
\be
\del_\Ac = \del_z + \del_{\bar{z}},
\qquad
\del_\Bc = \tau\del_z + \bar{\tau}\del_{\bar{z}}.
\ee
In this notation, 
we can define the winding numbers $\om_1, \om_2$ of the scalar $\phi$
along the $\Ac$-cycle and $\Bc$-cycle
by the expression
\be
(d\phi)^{\z} = 2\pi \om_1 E_\Ac + 2\pi \om_2 E_\Bc 
\qquad
(\om_1, \om_2 \in \mathbb{Z}).
\ee
For convenience we also list here the full expression of $\phi$
\be
\phi(z, \zb) = \phi_0 + 2\pi \om_1 \s_\Ac + 2\pi \om_2 \s_\Bc
+ \tilde{\phi},
\label{phi-expand}
\ee
where $\tilde{\phi}$ is the non-zero mode 
\be
\tilde \phi = \sum_{m, n \in \mathbb{Z}} \Phi_{mn}b_{mn},
\ee
and $b_{mn}$ is the basis of Fourier modes defined in (\ref{bmn}).

The original FJ action
\be
S_0 =
\frac{1}{4 \pi g^2}\int dz d\zb \; (\del_\Ac\phi) (\del_{\zb}\phi)
\label{FJ-S0}
\ee
has the equation of motion
\be
\del_\Ac \del_{\zb}\phi = 0.
\label{FJ-EOM}
\ee
This is not equivalent to the self-duality condition
\be
\del_{\zb}\phi = 0
\label{FJ-SD}
\ee 
for a chiral boson,
but it is invariant under the transformation
\be
\phi \rightarrow \phi' = \phi + F(\sigma_\Bc).
\label{FJ-gauge}
\ee
Imposing this transformation as a gauge symmetry of the theory,
one can easily verify that 
the equation of motion (\ref{FJ-EOM}) is gauge-equivalent to 
the chiral boson condition (\ref{FJ-SD}).
The existence of an additional gauge symmetry 
is a salient feature of all Lagrangian formulations for chiral bosons, 
and it distinguishes the previous two approaches using non-chiral boson action 
from the approach introduced below.

Since the gauge transformation parameter $F(\sigma_\Bc)$ depends only on a single variable, 
the gauge-fixing condition has to be chosen correspondingly.
Furthermore, 
the action $S_0$ (\ref{FJ-S0}) is gauge-invariant 
only if $F(\s_\Bc)$ has its winding number equal to zero.
Hence the gauge transformation cannot change the winding number of $\phi$ 
along the $\Bc$-cycle.
That is, the winding number $\om_2$ is gauge-invariant.

Consider the gauge-fixing condition
\be
\oint_{C_\Ac(\sigma_\Bc)} E_\Ac \del_\Bc \phi(z, \zb) = 2\pi \om_2 = \mbox{constant}
\qquad (\forall \s_{\Bc}),
\label{gauge-fixing}
\ee
where the cycle $C_\Ac(c)$ is an $\Ac$-cycle defined by $\sigma_\Bc = c$.
The gauge-fixing condition is an integral over a cycle such that 
it is a one-variable function because the gauge transformation (\ref{FJ-gauge})
is parametrized by a single-variable function $F(\sigma_\Bc)$.
The gauge-fixing condition can be imposed by introducing a Lagrange multiplier 
$\lam(\sigma_\Bc)$ which integrates to zero on $C_\Bc$ 
and an additional term 
\be
i\int_{C_\Bc} E_\Bc \lam(\sigma_\Bc) \left[\int_{C_\Ac(\sigma_\Bc)} E_\Ac \del_\Bc \phi(z, \zb)\right]
= i\int E_\Bc\wedge E_\Ac \; \lam(\sigma_\Bc) \del_\Bc\phi(z, \zb)
\label{gauge-fixing-Lagrange}
\ee
in the action.

Apparently the ghost fields associated with this gauge-fixing condition
should live on a one-dimensional space and 
they are decoupled from the chiral boson.
They contribute to an overall constant 
(independent of the source field)
to the partition function.
We will ignore the ghosts for simplicity here and 
consider the ghost action in Sec. \ref{s:BRS}.

As we noted earlier, 
the winding number $\om_2$ is gauge-invariant.
We should remove it since the chiral boson does not 
have this physical degrees of freedom.
(This will be justified below in Sec. \ref{Ham}.)
To impose a constraint to remove $\om_2$,
we introduce a (constant) Lagrange multiplier $a \in \mathbb{R}$
and an additional term in the action
\be
S_a[a] = i2\pi \om_2 a.
\label{Sa}
\ee
Due to the presence of this term, 
in the following we will set 
\be
\om_2 = 0.
\label{omega2=0}
\ee
Incidentally, this term can be combined with 
the gauge-fixing term (\ref{gauge-fixing-Lagrange}) 
into a single term in the action
\be
S_B =
i\int E_\Bc\wedge E_\Ac \; B(\sigma_\Bc) \del_\Bc\phi(z, \zb),
\label{lamdelBphi}
\ee
where the Lagrange multiplier $B(\sigma_\Bc)$ is no longer 
restricted to be free of the constant mode.

The gauge-fixing condition still admits a residual gauge symmetry
which is the shift of $\phi$ by a constant
\be\label{phi-shift}
\phi \rightarrow \phi' = \phi + c.
\ee
This gauge symmetry allows us to gauge-fix the constant mode 
in $\phi$ to be zero.  We will observe in Sec. \ref{Ham} that
the physical state is invariant under such shift.

Now we consider the source term.
Due to this additional gauge symmetry (\ref{FJ-gauge}),
only the gauge-invariant field strength 
$\del_\Ac \phi$ should be coupled to a physical source field $A$.
We define the source term of the action as
\be
S_A[A] = \frac{1}{2\pi g} \int dz d\zb \; (\del_\Ac\phi) A.
\label{SA}
\ee
This is the only term in the action where the source field $A$ appears.
In terms of the variable $y$
\be
y = \frac{i \tau_2}{\pi} A
\ee
that was also defined in other approaches,
the zero mode contribution to $S_A$ is
\be\label{e:SAz}
S_A^{\z} = -i\frac{2\pi}{g} \om_1 y.
\ee
Since the zero mode of $y$ only couples to the winding number $\om_1$ through this term, 
we see that a shift of $y$ by 
\be
y \rightarrow y' = y + g n
\label{gauge-transf-y}
\ee
for an integer $n$
induces a shift of the action by $i2\pi n$.
Thus the partition function $Z_0[y]$ is periodic,
\be
Z_0[y+g] = Z_0[y].
\label{period}
\ee

So far we have introduced the original FJ action 
and the source term, 
which are already present for a base space with trivial topology.
When the base space has nontrivial topology,
there is an ambiguity in the choice of topological terms, 
which do not change the local equation of motion for the scalar $\phi$.
In principle, 
to explore all models of chiral boson with its local physics 
described by the FJ action, 
we should consider all topological terms consistent with 
all symmetries of the theory.

The topological term $\int d\phi \wedge d\phi$ is trivial, 
and the only nontrivial topological term
$\int d\phi \wedge \kappa$ for some constant one-form $\kappa$
is equivalent to a shift of the source field $A$.

To consider more general Lagrangians with additional topological terms, 
we introduce an auxiliary scalar field $\Psi$.
Then we have the following additional topological terms 
\be
\int d\phi \wedge d\Psi
\qquad \mbox{and} \qquad
\int \Gamma \wedge d\Psi
\label{top-terms}
\ee
for some constant one-form $\Gamma$.
One immediately realizes that 
the non-zero-mode part of the auxiliary field $\Psi$ is irrelevant 
because it only appears in total derivatives.
The winding modes of $\Psi$, 
which are coupled to $\omega_1$,
have the effect of introducing a periodicity to the source field.

Since the source field already has a periodicity (\ref{period}),
the most general consistent possibility is to impose a (possibly twisted) periodicity 
which is a fraction of the period already present.
Therefore,
we should focus on the case when 
the effect of the first term in (\ref{top-terms})
is to take a quotient of the periodicity (\ref{period}) by $\mathbb{Z}_{k_1}$ 
for a positive integer $k_1$.
In other words, 
one imposes a twisted periodic boundary condition 
on a redefined partition function
\footnote{
This condition (\ref{quotient}) implies the periodicity (\ref{period}).
}
\be
Z_{0}^{new}[y+g/k_1] = e^{i2\pi n_1/k_1} Z_{0}^{new}[y]
\label{quotient}
\ee
for a fractional shift
\be
y \rightarrow y' = y + g/k_1.
\ee
Here the parameter $n_1$ is an integer (defined modulo $k_1$)
so that (\ref{period}) still holds.
This new theory is specified by two more integral parameters $k_1, n_1$.

Given a function $Z_0$ with the boundary condition (\ref{period}),
one can use the method of images to construct a new function $Z^{new}$ with 
the new boundary condition (\ref{quotient}).
Simply define the new partition function as
\be
Z_0^{new}[y] = \sum_{n=0}^{k_1-1} e^{- i2\pi n n_1/k_1} Z_0[y + gn/k_1],
\label{method-images}
\ee
then (\ref{quotient}) is satisfied.
Extending the range of values of the parameter $n$ to all integers,
and treating the symmetry 
\be
n \rightarrow n + r k_1 
\qquad 
\forall r \in \mathbb{Z}
\label{nrk1}
\ee
as a gauge symmetry,
the new partition function can be rewritten as
\be
Z_0^{new}[y] = 
\frac{1}{(\sum_{r\in\mathbb{Z}}1)}\sum_{n\in\mathbb{Z}} e^{- i2\pi n n_1/k_1} Z_0[y + gn/k_1],
\label{method-images-1}
\ee
where $(\sum_{r\in\mathbb{Z}}1)$ is the volume of the gauge symmetry (\ref{nrk1})
one should remove from the path integral.
The shift in $y$ by $gn/k_1$ on the right hand side of (\ref{method-images-1})
can be implemented by introducing an auxiliary ``field''
\be
\psi = 
\frac{\pi}{i\tau_2}\frac{n}{k_1}
\qquad (n \in \mathbb{Z})
\label{quantize-psi}
\ee
and replacing $A$ by $A + g\psi$
(or equivalently $y$ by $y + gn/k_1$) in the source term.
$\psi$ is called a ``field'' because
the variable $n$ should be summed over in the path integral 
like a physical variable.
Later we will show that $\psi$ is embedded in the auxiliary field $\Psi$
introduced in (\ref{top-terms}).
In addition, the phase factor $e^{-i2\pi nn_1/k_1}$ 
on the right hand side of (\ref{method-images-1}) can be implemented 
by adding another term
\be
S_{\gamma}[\psi] = \frac{1}{2\pi}\int dzd\zb \; \gamma \psi
\ee
in the action,
where
\be
\gamma = -2\pi n_1.
\label{gamma}
\ee
This term $S_{\gamma}$ will be rewritten 
in the form of the second term in (\ref{top-terms}).

Summarizing, 
we mentioned the modification of the action by shifting 
the source field by a background field, 
and by $S_{\gamma}$ to take the quotient of the periodicity (\ref{period}).
The total action 
is thus
\be
S_{\phi} = S_0 + S_B + S_A[A+A_0+g\psi] + S_{\gamma}[\psi],
\label{total-S}
\ee
where $A_0 
= \frac{\pi}{i\tau_2}\frac{g\beta}{k_1}$.
$\beta$ will be the convenient variable parametrizing the background source.

The guiding principle for the modifications above to the action $S_0 + S_A[A]$ 
is the quotient condition (\ref{quotient}).
To impose this condition we introduced an auxiliary field $\psi$
in order to apply the method of images.
We also consider additional modification to the action 
without breaking any symmetry or the boundary condition (\ref{quotient}),
such as introducing a constant shift $A_0$ for the source field $A$.
A change of the definition of the auxiliary field $\psi$ or the parameter $\gamma$ 
would typically lead to a trivial partition function.
For example, if we have defined $\psi$ to take continuous, rather than discrete, values,
integrating over $\psi$ in the path integral 
would give a partition function independent of $A^{\z}$.

For a generic interacting chiral boson Lagrangian, 
which may or may not be Lorentz-invariant,
we propose to introduce the same correction terms $S_A[A+A_0+g\psi] + S_{\gamma}$ 
to define its partition function.
More generally, one should consider all possible topological terms, 
as one would normally do when considering a class of theories.
For the example at hand, there are no other nontrivial topological terms to consider.

Let us now show that
the correction terms to the original FJ action,
including the source term
($S_0 + S_A[A+A_0]$) in (\ref{total-S}),
can be written as topological terms of the form (\ref{top-terms}).
Indeed, consider
\be
S_{\Psi}[\Psi] = \frac{i}{2\pi k_1} \int (d\phi + \Gamma) \wedge d\Psi.
\label{S1}
\ee
As the integrand is a total derivative,
only the zero modes of each differential contribute to the action.
The auxiliary field $\Psi$ is a scalar field with the same compactified target space 
as that of $\phi$,
\be
\Psi \sim \Psi + 2\pi N
\qquad
(N \in \mathbb{Z}).
\label{Psi-BC}
\ee
Its winding modes are given by
\be
(d\Psi)^{\z} = 2\pi m E_\Ac + 2\pi n E_\Bc
\qquad 
(m, n \in \mathbb{Z}),
\label{dPsi}
\ee
where $m, n$ are the winding numbers of $\Psi$.
While the winding number $m$ is decoupled because $\om_2 = 0$,
the winding number $n$ is precisely the same variable 
we used to define $\psi$ (\ref{quantize-psi}).
As we promised earlier, 
the auxiliary field $\psi$ is embedded in an auxiliary scalar field $\Psi$ 
with non-zero modes,
and the correction terms $S_A[g\psi]$ and $S_\gamma[\psi]$
can be expressed as manifestly topological terms.

In (\ref{S1}),
the constant one-form $\Gamma$ 
is defined by two constant parameters $\gamma$ and $\zeta$ as
\be
\Gamma = \gamma E_\Ac + \zeta E_\Bc,
\ee
where $\gamma$ is defined in (\ref{gamma}), 
and $\zeta$ is coupled to the winding number $m$ in $d\Psi$ (\ref{dPsi}).
As $m$ is decoupled from $\phi$, 
the contribution of $\zeta$ is merely an overall factor of the partition function.
We will simply ignore the parameter $\zeta$ in our calculation 
of the partition function.
\footnote{
In order for the partition function to be non-vanishing,
$\zeta$ has to be multiples of $2\pi k_1$.
However, as long as the partition function is non-vanishing, 
all choices of $\zeta$ are equivalent.
}

The effect of introducing a background $A_0$ to the source field
can be replaced by twisting the boundary condition of $\Psi$ (\ref{Psi-BC}) as
\be
\Psi \sim \Psi + 2\pi (N+\beta)
\qquad (N \in \mathbb{Z})
\ee
for a real parameter $\beta$ (defined modulo $1$).
This is equivalent to the well understood connection between Wilson loop 
and twisted periodic boundary condition of charged fields.

Finally, the complete action is
\bea
S_{\phi} &=& S_0 + S_A[A+A_0] + S_{\Psi} + S_B
\nn \\
&=&
\frac{1}{4 \pi g^2}\int dz d\zb (\del_z+\del_{\bar{z}})\phi \del_{\zb}\phi 
+ \frac{1}{2\pi g}\int dz d\zb \; (\del_z+\del_{\zb})\phi (A+A_0) 
\nn \\
&&
+ \frac{i}{2\pi k_1} \int (d\phi + \Gamma) \wedge d\Psi
+ i\int E_\Bc\wedge E_\Ac \; B(\sigma_\Bc) \del_\Bc\phi(z, \zb).
\label{totalS}
\eea
The first term is the original FJ action. 
The second term is the source term with a background $A_0$.
The third term is a topological term with an auxiliary field $\Psi$ 
introduced to impose the quotient condition (\ref{quotient}).
Since the non-zero modes of $\Psi$ vanish completely in the action,
only the winding modes of $\Psi$ are relevant.
Hence $\Psi$ should not be viewed as an ordinary 2-dimensional field.
The last term is essentially the gauge-fixing term, 
so it would be changed if we choose a gauge-fixing condition 
different from (\ref{gauge-fixing}).

\subsection{$Z_0[A^{\z}]$}

For the FJ action modified by topological terms (\ref{total-S}), 
the partition function for the zero mode of $A$ is easily calculated.
It is
\be
Z_0[A^{\z}] =
{\cal N} \sum_{\om_1, n} 
e^{-S_{\phi}^{\z}} 
= 
\vth\col{n_1/k_1}{\beta}\left(\frac{k_1}{g}y; \frac{k_1^2}{g^2}\tau\right),
\label{e:gtheta}
\ee
where 
$S_{\phi}^{\z}$ 
denotes the contribution of winding modes to the action
(\ref{totalS}).
The normalization factor ${\cal N}$ involves dividing 
the path integral by the volume of all gauge symmetries.
Effectively the index $n$ is summed from $0$ to $k_1-1$.
To derive this result we changed variables by
\be
\om_1 = k_1 p + u, 
\qquad
n = k_1 q + v,
\ee
where $p, q \in \mathbb{Z}$ and $u, v = 0, 1, 2, \cdots, (k_1-1)$.
The partition function involves an overall infinite constant factor 
${\cal N}^{-1} = k_1\left(\sum_{q}\right) $, 
reflecting the fact that 
there is a gauge symmetry of infinite order 
corresponding to the shift of the auxiliary field $\psi$
by multiples of $g\pi/(i\tau_2)$, 
which is equivalent to the shift of the source field $A$
according to (\ref{gauge-transf-y}).

Apparently the coupling constant that would reproduce 
the result of holomorphic decomposition (\ref{chiralZ0}) is
\be
g = \sqrt{k_1 k_2}.
\ee
For the non-chiral theory, the coupling constant $g$ can be 
either $\sqrt{k_1 k_2}$ or $2/\sqrt{k_1 k_2}$,
which are related by T-duality, 
and therefore both choices are allowed (and equivalent) 
in the approach of holomorphic decomposition.
For a chiral boson theory there is no T-duality,
and one may wonder which choice survives
when only the holomorphic part of the source $A_{\zb}$ is coupled to the chiral boson.
In the approach of holomorphic line bundle,
we only see the possibility of $g = 2/\sqrt{k_1 k_2}$,
while the other possibility shows up here.
Despite the superficial difference, 
the partition function is given by $\vth$ functions
for all three approaches.

It is possible to lift the rational characteristic parameter $n_1/k_1$
of the $\vth$ function to a generic real number by 
introducing twisted boundary conditions on $\phi$,
or equivalently by introducing constant connections on the torus.
In this paper we have focused our attention on the periodic boundary condition 
for the chiral boson $\phi$ without twisting for the sake of comparison 
among different approaches.
If we take into consideration the twisting of the boundary condition
\be
\phi(z+m+n\tau, \zb+m+n\bar{\tau}) = \phi(z, \zb) + 2\pi[m(\om_1+\a) + n(\om_2+\b)]
\ee
defined by two real parameters $\a, \b \in \mathbb{R}/\mathbb{Z}$,
the characteristic parameters of the $\vth$ function would be shifted 
by $\a$ and $\b$.
The twisting of the boundary condition is equivalent to 
turning on a constant connection 1-form $C = 2\pi (\a E_\Ac + \b E_\Bc)$
so that all the derivatives in the action are replaced by 
the covariant derivatives $D_i = \del_i + C_i$.

Apparently, 
this approach provides the most straightforward calculation
of the zero-mode partition function 
among all three approaches discussed in this paper.

\subsection{$\tilde{Z}[\tilde{A}]$}

Now we compute the partition function for the non-zero modes
\be
\tilde{Z} = \int D\tilde{\phi} \; e^{-\tilde{S}_{\phi}},
\ee
where
\be
\tilde{S}_{\phi} = \frac{1}{4\pi g^2}\int dzd\zb \;\left(
(\del_z+\del_{\zb})\tilde{\phi}\del_{\zb}\tilde{\phi} 
+ 2g(\del_z+\del_{\zb})\tilde{\phi} \tilde{A}\right),
\ee
up to the gauge-fixing terms 
which will be ignored for the calculation below.

The partition function satisfies the differential equation
\be
\del_{\zb}\frac{\d}{\d \tilde{A}} \tilde{Z}
= \frac{1}{2\pi}(\del_z+\del_{\zb}) \tilde{A} \tilde{Z},
\ee
where we used the identity 
\be
\langle (\del_z+\del_{\zb})\del_{\zb}\tilde{\phi} + g (\del_z+\del_{\zb})\tilde{A} \rangle = 0.
\ee

Let 
\be
\tilde{Z} = \tilde{N} e^{\int\int dzd\zb dz'd\zb' \; K(z-z',\zb-\zb') \tilde{A}(z,\zb) \tilde{A}(z',\zb')},
\ee
where $\tilde{N}$ is independent of $\tilde{A}$,
we find that $K$ needs to satisfy the equation
\be
\del_{\zb}K(z-z', \zb - \zb') = -\frac{1}{4\pi} (\del_z+\del_{\zb})\d^{(2)}(z-z', \zb - \zb').
\ee
Assuming that $G(z-z', \zb - \zb')$ is the Green's function satisfying (\ref{Laplace}),
we have 
\bea
K(z-z', \zb-\zb') &=& 
- \frac{1}{4\pi}(\del_z+\del_{\zb})\del_z G(z-z', \zb-\zb') 
+ K_0(z-z') 
\nn \\
&=&
- \frac{1}{4\pi}\del_z^2 G(z-z', \zb-\zb') 
- \frac{1}{4\pi}\d^{(2)}(z-z'; \zb-\zb')
+ K_0(z-z'). 
\nn \\
\eea
Apart from the 2nd term in the last expression, 
this solution of $K$ is exactly the same as the kernel (\ref{KG})
derived above in the approach of holomorphic line bundle.
The extra term contributes to an extra factor of
\be
e^{-\frac{1}{4\pi}\int dz d\zb \; \tilde{A}^2(z, \zb)}
\ee
in $\tilde{Z}$.
This is a self-contraction contribution that is removed by 
normal ordering.
Therefore, for the correlation functions of normal ordered operators,
the path integral of FJ action is in perfect agreement with 
the other approaches using the action of a non-chiral boson.
The equivalence with the result of $\tilde{Z}$ (\ref{ZKK-HNS}) 
for the holomorphic decomposition
can also be shown via mode expansions.

\section{FJ action and BRST symmetry}
\label{s:BRS}

\subsection{Gauge fixing and BRST symmetry}

In the previous section, we introduced a gauge-fixing condition
(\ref{gauge-fixing-Lagrange}). 
An important feature of such gauge fixing is that the
Lagrange multiplier $\lambda$ depends only on a single coordinate $\sigma_B$ on the torus, 
while in the usual case, the gauge parameter
depends on every coordinate on the space-time where gauge theory is defined.
Since such gauge symmetry is typical in the Lagrangian formulation of self-dual fields,
we give some detail of BRST quantization although most of the analysis
is elementary and straightforward.

Here we consider a general situation where the world-sheet topology may not
be restricted to torus.  Also we return to the Lorentzian space-time coordinates
$(\sigma_0, \sigma_1)$, instead of the Euclidean ones $(z,\bar z)$
since for the canonical quantization they are more convenient.
We require that the spatial coordinate $\sigma_1$ to be periodic $\sigma_1\sim \sigma_1+1$.

The FJ lagrangian (\ref{FJ-S0}) is rewritten with coordinates 
$\s_0, \s_1$
as
\ba\label{FJ-GF}
S_0=\frac1{4\pi g^2} 
\int d\s_0 d\sigma_1\partial_1\phi(\partial_0-\partial_1)\phi\,.
\ea
As we remarked, the action has a gauge symmetry (\ref{FJ-gauge})
which is written in terms of Lorentzian coordinates as
\be\label{FJ-gauge1}
\delta\phi = F(\s_0)\,.
\ee
The gauge-fixing term of the Lagrangian (\ref{gauge-fixing-Lagrange}) together with
the Lagrange multiplier field $B$ are introduced to fix this symmetry.
As the gauge parameter $F(\s_0)$ depends only on $\s_0$, so does
$ B $,
\ba
S_B = -\int  d\s_0 d\sigma_1
B(\s_0)(\partial_0 +h\partial_1)\phi(\s_0,\sigma_1)
\label{SGF},
\ea
where $h$ is an arbitrary real parameter. We note that
$\partial_B=\tau_1\partial_1+\tau_2\partial_2$ in 
 (\ref{gauge-fixing-Lagrange}) is replaced by
$\tau_0\partial_0+\tau_1\partial_1$,
where $\tau_0 = i\tau_2$.
However,
since we do not assume particular shape for the world-sheet,
we replace 
$\tau_1/\tau_0$ 
by an arbitrary parameter $h$.
The gauge condition by integrating out $B$,
\ba
\int d\sigma_1 (\partial_0 +h\partial_1)\phi(\s_0,\sigma_1)=0\,,
\label{e2}
\ea
certainly removes the gauge freedom (\ref{FJ-gauge1})
for any $h$.  There are a variety of possibilities for the
gauge-fixing conditions which may be used to fix  (\ref{FJ-gauge1}).
In this paper we restrict ourselves to such conditions
since the analysis seems to be the simplest.
We introduce further the Faddeev-Popov ghosts $b(\s_0), c(\s_0)$
with the action
\be
S_{FP} = -i \int d\s_0 b(\s_0)\partial_0 c(\s_0)
\label{SFP}\, ,
\ee
which was ignored in the previous section.
In addition, there are the source term $S_A[A+A_0]$ (\ref{SA}) and 
the topological term $S_{\Psi}$ (\ref{S1}) in the total action (\ref{totalS}).
These may be included when the world sheet is torus.
We drop these terms for a moment
and will be taken into account later in the computation of partition function.

The total action $S=S_0+S_B+S_{FP}$ has BRST symmetry
\ba
\delta\phi(\s_0,\sigma_1) = c(\s_0), 
\qquad
\delta B(\s_0) = 0, 
\qquad
\delta b(\s_0) = i B(\s_0), 
\qquad
\delta c(\s_0) = 0\,.
\ea

In the source-free theory ($A+A_0 = 0$),
the equation of motion which is derived from the total action is
\ba
&& \partial_1(\partial_0 -\partial_1)\phi= 2\pi g^2\partial_0 B(\s_0),
\label{e1}\\
&& \partial_0 c=\partial_0 b=0
\label{e3}\,,
\ea
together with (\ref{e2}).
We note that the first equation (\ref{e1}) does not look like
the self-dual equation $(\partial_0 -\partial_1)\phi=0$.
However, solving the equation of motion shows indeed they are equivalent.

Let us define the periodicity of $\phi$ by,
\ba\label{period-phi}
\phi(\sigma_1+1)=\phi(\sigma_1)+2\pi \omega, 
\qquad 
\omega\in \mathbb{Z}\,.
\ea
We introduce a Fourier transformation of $\phi$
compatible with periodicity (\ref{period-phi}),
\ba\label{f}
\phi(\s_0,\sigma_1) = x(\s_0)+2\pi \om\sigma_1 
+g \sum_{m\neq 0}\frac{i}{m} a_m(\s_0) e^{-2\pi im\sigma_1}\,.
\ea
Plugging it into (\ref{e1}) gives
\ba
2\pi g \sum_{m\neq 0} (\dot a_m+2\pi im a_m) e^{-2\pi im\sigma_1}= 2\pi g^2 \partial_0 B\,.
\ea
By comparing each Fourier mode, we obtain
\ba
\dot a_m +2\pi im a_m=0,
\qquad
\partial_0 B=0,
\ea
which imply
\ba
a_m(\s_0) = c_m e^{-2\pi im\s_0}.
\ea
The constraint (\ref{e2}) gives
$
\dot x+2\pi \omega h=0\,.
$
It gives the solution of the equation of motion in the form
\ba\label{p1}
\phi(\s_0,\sigma_1) = x(0)+2\pi \omega(\sigma_1-h\s_0) 
+g \sum_{m\neq 0}\frac{i}{m} c_m e^{-2\pi im(\s_0+\sigma_1)}\,.
\ea
We note that the zero-mode has wrong dependence on $\s_0,\sigma_1$
(i.e. $\del_0 \phi \neq \del_1 \phi$)
unless $h=-1$. 
On the other hand, 
the oscillator part has the correct form.

To make comparison with the previous section, 
one may replace 
$h=\tau_1/\tau_0$
and take Wick rotation. The zero-mode part becomes
$\phi\sim 2\pi \om \sigma_\mathcal{A}$.  It describes the winding of $\phi$
around the cycle $\mathcal{C}_\mathcal{A}$ which correctly
reproduces the set-up in the previous section, namely
$(\omega_1,\omega_2)=(\omega,0)$.

\subsection{Canonical quantization and physical states}
\label{Ham}
Let us go further to analyze the Hilbert space of the system 
by canonical quantization.
Since the action is first order in time derivative, we need to introduce
the second class constraint.
Such issue was already solved in \cite{Sonnenschein:1988ug}.
Here, instead of introducing constraint, we restrict
the set of the coordinates to remove the redundancy of variables.
Such a quantization scheme is close to the treatment of original literature
\cite{Floreanini:1987as}.

We put the Fourier expansion (\ref{f}) into the action
and arrive at
\ba
S_0 &=& \frac{1}{2g^2}
\int d\s_0 \omega(\dot x-2\pi \omega)
+\int d\s_0 \sum_{m>0} a_{-m} \frac{i}{m}(\dot a_m +2\pi i m a_m),
\\
S_B &=&-\int d\s_0 B(\s_0)(\dot x+2\pi \omega h)\,.
\ea
The second term of $S_0$ implies that the momentum for $a_m$ is 
$\frac{i}{m} a_{-m}$ ($m>0$) and that for $x$ is $-B$.
One may set the (equal time) canonical commutation relations as
\ba
[a_n, a_m]= n\delta_{n+m},
\qquad
[x,B]=-i,\qquad \{b,c\}=1\,.
\ea
(Others like $[a_n, B]$ etc. vanish.)

The Hamiltonian reads
\ba
H=2\pi \left( \sum_{n>0} a_{-n} a_n +\frac{\omega^2}{2 g^2} + \om h B \right)\,.
\ea
The momentum operator becomes
\ba
P&=& 
\int d\sigma_1 \frac{\d S}{\d\dot \phi}\partial_1\phi
=H - 2 \pi (1+h) \omega B\,.
\ea
The second term in this expression looks strange and it would imply
that the self-duality condition $H=P$ is broken when $h\neq -1$.
However, as we see below, the extra term vanishes when
applied to the physical state.  So the self-duality holds in the above sense for any $h$
and does not depend on the choice of gauge fixing.

The nilpotent BRST operator is defined as
\ba
Q= iB c\,.
\ea
One may easily confirm that
\ba
Q^2=0,
\qquad 
[Q,\phi]= c,
\qquad
\left\{Q,b\right\}= i B,
\qquad 
\left\{Q,c\right\}=[Q,B]=0\,,
\ea
and that the BRST transformation is generated by $Q$,
$\delta\Psi=[Q,\Psi\}\,.$
The physical state condition is given in terms of $Q$ as
\ba\label{brs1}
Q|\Psi\rangle =0,
\qquad 
|\Psi\rangle \neq Q|\chi\rangle\,.
\ea
Focusing on the zero-mode part of the physical states,
we have
\ba
|\Psi\rangle =\psi_0(x)|0, \om\rangle +\psi_1(x) |1, \om\rangle,
\qquad
|\chi\rangle =\chi_0(x)|0, \om\rangle +\chi_1(x) |1, \om\rangle
\ea
with $c|1, \om\rangle=b|0, \om\rangle=0$, $|1, \om\rangle = c|0, \om\rangle$.
Eqs.(\ref{brs1}) are rewritten as
\ba
\partial_x \psi_0(x)=0,
\qquad 
\psi_1(x)\neq \partial_x \chi_0(x)\,.
\ea
It implies that the zero-mode part of the physical states is 
a superposition of $|0, \om\rangle$ and $|1, \om\rangle$ independent of $x$.
To avoid redundancy, we impose the constraint 
\be
b|\Psi\rangle = 0
\ee
on physical states, 
so that the only nontrivial element in BRS cohomology 
in the zero-mode sector is
\ba
|\Psi\rangle = |0, \om\rangle \, .
\ea
Namely the dependence on $x$ for the physical state is gone.
In other words, $B|\mbox{phys}\rangle=0$,
which removes the undesirable term in $P$.  At the same time,
it gauge-fixes the gauge symmetry (\ref{phi-shift}) in the previous section.
The physical Hilbert space is spanned by
states generated by oscillators,
\ba
a_{-n_1}\cdots a_{-n_\ell}|0, \om\rangle\, .
\ea
We denote the Hilbert space spanned by these basis as $\mathcal{H}$.

We introduce an operator $\Omega$ which takes the winding number
as its eigenvalue and commutes with the oscillators,
\ba
\Omega|0,\omega\rangle =\omega |0,\omega\rangle,
\qquad
[\Omega, a_n]=0\,,
\ea
and it also commutes with $H$ and $P$.
In terms of the field $\phi$, it can be written as
\ba
\Omega=\frac{1}{2\pi}\int d\sigma_1 \partial_1 \phi\,.
\ea
In the previous section, we add a term
(\ref{e:SAz}) to the action which expresses the coupling of the zero mode of $A$. 
With the identification $\omega_1=\omega$,
it gives a twist in the definition of the partition function,
\ba
Z&=&\mbox{Tr}_\mathcal{H} (e^{i(\tau_1 P+\tau_0 H) +2\pi i (y/g) \Omega})
=\mbox{Tr}_{\mathcal{H}} (q^{\frac{H+P}{4\pi}} \bar q^{\frac{H-P}{4\pi}} e^{2\pi i (y/g) \Omega})\,,\\
q &=&e^{2\pi i(\tau_1+ \tau_0)}\,,\qquad \bar q =e^{-2\pi i(\tau_1- \tau_0)}\,.
\ea
One immediately finds
\ba
Z=\frac{\sum_\omega q^{\frac{\omega^2}{2g^2} } e^{2\pi i(y/g)\omega}}{\prod_{n=1}^\infty (1-q^n)}\,,
\ea
namely the partition function of the chiral boson on the torus.
The factor in the numerator is the theta function
$
\vartheta\left(\frac{y}{g}, \frac{\tau_1+\tau_0}{g^2}\right)\,.
$

In the previous section, we further defined a coupling to auxiliary field $\psi$
to make the partition function with refined periodicity (\ref{quotient}).
The procedure may be explained as follows.
In addition to the Hilbert space $\mathcal{H}$, we take another 
finite dimensional Hilbert space $\mathcal{H}_\psi$ for auxiliary field spanned 
by $|n\rangle_\psi$, ($n\in \mathbf{Z}_{k_1}$) and take a tensor product $\mathcal{H}\times \mathcal{H}_\psi$.
We assume $H|n\rangle_\psi=P|n\rangle_\psi =0$
and introduce an operator which distinguishes the basis,
\ba
U|n\rangle_\psi = e^{2\pi i n/k_1}|n\rangle_\psi\,.
\ea
The new partition function by adding the topological terms is obtained in the operator formalism as
the trace over $\mathcal{H}\times \mathcal{H}_\psi$,
\ba
Z^{new} &=&\mbox{Tr}_{\mathcal{H}\times \mathcal{H}_\psi} 
(q^{\frac{H+P}{4\pi}} \bar q^{\frac{H-P}{4\pi}} e^{2\pi i (y/g) \Omega}U^{\Omega-n_1})
\,,
\ea
which reproduces the theta function with characteristics (\ref{e:gtheta}).


Before closing this section,
let us comment on the choice of the parameter $h$
in the Lagrangian formulation.
We take Wick rotation,
and recover the extra terms in the action.
In the notation in the previous section, the full action is
\bea
S &=& 
\frac{1}{4 \pi g^2}\int dz d\zb \; (\del_\Ac\phi) (\del_{\zb}\phi)
+ \frac{1}{2\pi g} \int dz d\zb \; (\del_\Ac\phi) (A+A_0+g\psi)
+ \frac{1}{2\pi}\int dzd\zb \; \gamma \psi
\nn \\
&&
- i\int E_\Ac\wedge E_\Bc \; B(\sigma_\Bc) \del_\Bc\phi(z, \zb)
- i \int E_\Ac\wedge E_\Bc \; b(\s_\Bc)\partial_\Bc c(\s_\Bc),
\label{S-full}
\eea
where we choose $h=\tau_1/\tau_0$.

It sounds rather strange that we need to take a special choice
for the gauge-fixing condition.  Indeed if we take general $h$,
the fourth term in (\ref{S-full}) takes the form
\ba
- i\int E_\Ac\wedge E_\Bc \; B(\sigma_\Bc) (\del_\Bc+h' \del_\Ac)\phi(z, \zb)
\ea
for $h'=h-\tau_1/\tau_0$.  
If we expand the zero mode of $\phi$ in the path integral 
(\ref{phi-expand}),
the integration over the zero-mode of $B$ gives
\ba
\omega_2+h' \omega_1=0\,.
\ea
In general for arbitrary $h$ and integers $\omega_1, \omega_2$, such equation 
has no nontrivial solution.  Therefore, in order to have nontrivial winding number
in the path integral, we need to set $h=\tau_1/\tau_0$. While the self-duality
holds for any $h$ in the canonical quantization, there seem to be an
obstruction in the path integral.

\section{Discussion}

In this work we re-examined existing approaches using non-chiral bosons 
to compute the partition function of a chiral boson, 
we also found a way to derive the same result from 
the Lagrangian of a chiral boson, 
by adding topological terms and an auxiliary field.
The well known result in the literature of partition function 
for the zero modes of the source field
is given by the $\vth$ functions
\be
\vth\col{0}{0}(y; \tau), \qquad
\vth\col{0}{\frac{1}{2}}(y; \tau), \qquad
\vth\col{\frac{1}{2}}{0}(y; \tau), \qquad
\vth\col{\frac{1}{2}}{\frac{1}{2}}(y; \tau).
\label{well-known}
\ee
We found the more general expression
\be
\vth\col{n_1/k_1}{\beta}\left(\frac{k_1}{g}y; \frac{k_1^2}{g^2}\tau\right),
\ee
which includes the special cases with $g = \sqrt{k_1 k_2}$ and
\be
\vth\col{\frac{n_1}{k_1}}{\frac{n_2}{k_2}}\left(\sqrt{\frac{k_1}{k_2}}y; \frac{k_1}{k_2}\tau\right)
\ee
derived using the method of holomorphic decomposition.
For $k_1 = k_2$,
the modular parameter $(k_1/k_2)\tau$ of the $\vth$ function 
is the same as that of the spacetime 2-torus $\Sigma$.
When one makes a modular transformation in spacetime,
the $\vth$ function changes by an overall phase factor, 
as a projective representation of the modular symmetry.
Even in this class we allow more general characteristic parameters than $0$ or $1/2$.

The issue with $k_1 \neq k_2$ is that the symmetry of modular transformations 
on the spacetime torus is partially broken.
However, part of the spacetime isometry is broken 
even for the well known cases (\ref{well-known}).
As an example, 
consider the parity symmetry $\sigma_1 \rightarrow - \sigma_1$
when $\tau_1 = 0$.
For instance, the expectation value
\be
\langle \del_1 \phi \rangle,
\ee
can be calculated from the equality
\be
\frac{\del}{\del A ^{(0)}}\left. Z(A^{\z})\right|_{A^{\z}=0}
\sim \int_{T^2} \langle \del_1 \phi \rangle.
\ee
For the well known cases (\ref{well-known}),
the zero-mode contribution to the partition function is $\vth(y|\tau)$, 
where $\vth$ represents one of the $\vth$ functions in (\ref{well-known}).
Hence we have
\be
\vth'(0|\tau) \sim \langle \del_1 \phi \rangle.
\ee
This means that the vacuum expectation value of winding numbers may be nonzero
if the corresponding $\vth$ function has a non-zero derivative at the origin.
This indeed happens when $\vth = \vth\col{1/2}{1/2}(y; \tau)$.
When $\tau_2 = 1$, 
there is an additional isometry of rotating the 2-torus by 90 degrees:
$z \rightarrow i z$.
In this case not only the parity symmetry is broken, 
but the symmetry of rotating by 90 degrees is also broken.
Therefore, the symmetry breaking for the isometry of a torus 
should be viewed as a generic feature of chiral boson theories,
and so we should not insist on preserving the full modular transformation 
as symmetries at the quantum level as the definition of 
a quantum chiral boson theory.

What we have learned from this study of the free chiral boson theory 
is that for a given chiral boson Lagrangian one should consider 
more general Lagrangians modified by topological terms and auxiliary fields,
without changing local equation of motion nor propagating modes.
While some of the choices of topological terms and auxiliary fields 
lead to a trivial partition function, 
others are well-defined theories.
We learned that suitable choices are guided by the periodicity in the source field,
and that a careful study of the canonical formulation can be helpful.
It will be interesting to apply this approach to 
interacting chiral boson theories 
for which its non-chiral version is not defined, 
so that the other two approaches 
(holomorphic decomposition and holomorphic line bundle)
are not applicable.

In this work we have restricted ourselves to the 2-torus 
as the base space of the chiral boson.
A complete understanding of the chiral boson in 2 dimensions 
should include Riemann surfaces of all topologies.
While chiral fermion is easier to understand, 
and it can be related to chiral boson via bosonization,
the bosonization on higher genus Riemann surfaces 
was already studied for 
chiral fermions
\cite{Verlinde:1986kw,Eguchi-Ooguri,AlvarezGaume:1987vm,IMO,Vafa,AMNVB}.
This provides a good starting point to understand chiral bosons 
on generic Riemann surfaces.
It will be interesting to construct a chiral boson Lagrangian 
with topological terms suitable for Riemann surfaces of higher genus.

Chiral bosons (or self-dual gauge fields) are 
the fundamental building blocks of bosonic fields, 
analogous to their fermionic counterparts.
While Weyl fermions play important roles in various models 
of high energy physics and condensed matter physics,
it should not be surprising if chiral bosons will play crucial roles 
in constructing physical models in the future.
Understanding the quantum theory of interacting chiral boson is thus 
an important fundamental issue in quantum field theory.
This work provides a first step towards conquering this challenge.

\section*{Acknowledgements}

The authors would like to thank 
Chong-Sun Chu and Takeo Inami
for their interest and discussions.
PMH is supported in part by
the National Science Council, Taiwan, R.O.C.
YM is partially supported by Grant-in-Aid
(KAKENHI \#25400246)
from MEXT Japan.

\appendix

\section*{Appendix}

\subsection*{$\vth$ function}

The modified $\vth$ function is defined by
\be
\vth\col{\a}{\b}(y; \tau)
= \sum_{n\in\mathbb{Z}}\exp\left[\pi i (n+\a)^2\tau\right]
\exp\left[2\pi i (n+\a)(y+\b)\right].
\ee
They satisfy the relation
\be
\label{thetaR}\vth\col{\alpha}{\beta}(y+\theta\tau + \phi; \tau) = 
e^{-2i\pi\theta(y+\beta+\theta\tau/2+\phi)}\vth\col{\alpha+\theta}{\beta+\phi}(y; \tau)
\ee
for $\phi,\theta\in \mathbb{R}$.
For the speical case $\theta=n,\phi=m\in\mathbb{Z}$,
\be
\label{thetaZ}\vth\col{\alpha}{\beta}(y + n\tau+m; \tau)
= e^{-i2\pi ny - i\pi n^2 \tau + i2\pi (m\alpha - n\beta)}\vth\col{\alpha}{\beta}(y; \tau).
\ee

Under a modular transformation, 
the $\vth$ function changes by a phase
\beqa
\vth\left[\alpha\atop{\beta}\right]
\bigg(y; \tau+1\bigg) &=&
e^{-i\pi\alpha(\alpha+1)}\vth\left[\alpha\atop{\beta+\alpha+\frac{1}{2}}\right](y; \tau),
\\
\vth\left[\alpha\atop{\beta}\right]
\bigg(\frac{y}{\tau}; -\frac{1}{\tau}\bigg) &=&
\sqrt{-i\tau}
e^{i\pi y^2/\tau+2\pi i \alpha\beta}\vth\left[\beta\atop{-\alpha}\right](y; \tau).
\label{mod-transf-th}
\eeqa

\subsection*{Green's function on torus}

The standard Green's function on a 2-torus is
\be
G(z-w, \zb-\bar{w}) = -\frac{1}{4\pi} \log \left|
\frac{\vth_1(z-w; \tau)}{\eta(\tau)} \right|
+\frac{1}{2}\frac{(\mbox{Im}(z-w))^2}{\mbox{Im}\tau}.
\label{Green}
\ee
It satisfies the equation
\be
\del_z\del_{\zb}G(z-w, \zb-\bar{w}) = \d^{(2)}(z-w, \zb-\bar{w})
- \frac{1}{\mbox{Im}\tau},
\label{EOM-Green}
\ee
where we needed to add a constant on the right hand side
because the 2-torus is a compact space.
The Green's function is ``doubly periodic''.
That is,
$G(z + m + n\tau, \bar{z} + m + n\bar{\tau}) = G(z, \bar{z})$
for all $m, n \in \mathbb{Z}$.

In the above, $\eta$ is the Dedekind $\eta$ function. 
It can be defined as
\be
\eta(\tau) = e^{i\pi\tau/12} \prod_{n=1}^{\infty}(1-q^n),
\ee
where 
\be
q = e^{2i\pi\tau}.
\ee
The factor of Dedekind $\eta$ function in (\ref{Green}) 
only contributes an additive constant to the Green's function,
and does not affect the equation (\ref{EOM-Green}) at all.
We can thus replace it by any other factor 
without changing the kernel $K$ (\ref{KG}) for the chiral boson partition function.

\end{CJK} 
\end{document}